\documentclass[preprint,12pt]{elsarticle}

\usepackage{amssymb}
\usepackage{amsmath}

\newcommand{\be}{\begin{equation}}
\newcommand{\ee}{\end{equation}}
\newcommand{\ba}{\begin{eqnarray}}
\newcommand{\ea}{\end{eqnarray}}
\newcommand{\nn}{\nonumber\\}
\begin{document}

\begin{frontmatter}



\title{Causality and Stability in Relativistic
Hydrodynamics} 


\author{Sukanya Mitra}
\ead{sukanya.mitra@niser.ac.in}
\address{School of Physical Sciences, National Institute of Science Education and Research, An OCC of Homi Bhabha National Institute, Jatni-752050, India.}


\begin{abstract}
The causality and stability of a relativistic hydrodynamic theory is shown to require a consensus between, either (i) newer degrees of freedom apart from the fundamental ﬂuid fields, or (ii) a general hydrodynamic frame other than the Landau or Eckart compromising the field's first principle definition, unless the non-equilibrium derivative correction goes to inﬁnity. Any finitely truncated derivative correction (no matter how high it is) is shown to lead to an acausal theory, unless the corrections are infinitely summed up to include all orders. From an underlying microscopic theory, an exact form of relativistic hydrodynamics has been derived which establishes that the resummation of all order temporal derivatives is essential for causality, which finally ‘integrated in’ as newer degrees of freedom.
\end{abstract}


\end{frontmatter}

\section{Introduction}
The benchmark criteria that make a hydrodynamic theory physically acceptable are, (a) causal wave propagation and, (b) stability against small perturbations. Apart from that, a hydrodynamic theory prefers certain desirable features. First, (i) they are deﬁned only in terms of the fluid
dynamical variables, such as temperature ($T$), ﬂuid velocity ($u^{\mu}$), and chemical potential ($\mu$), which are related to some conserved quantities
characterizing the equilibrium. Second, (ii) even  out of equilibrium the ﬂuid ﬁelds have well deﬁned first principle definition fixed by some hydrodynamic frames like Landau or Eckart's, in which the out of equilibrium fields are gauged to their equilibrium values to have an unambiguous definition. The last one is that (iii) the theory must contain a ﬁnite number of derivatives over ﬂuid variables.

In reality, it is observed that these three features are simultaneously never achieved unless the causality (and stability) of the theory is compromised. The Navier-Stokes (NS) theory which is a finitely truncated ﬁrst-order theory defined at both Landau and Eckart frames \cite{LandauEckart}, ends up with superluminal signal propagation and thermodynamic instability. The higher (but ﬁnite) order derivative theories like the Muller-Israel-Stewart (MIS) theory \cite{MIS} required new degrees of freedom apart from $T,u^{\mu},\mu$ to maintain causality \cite{Hiscock}. The recently proposed BDNK theory \cite{BDNK} presents a causal-stable ﬂuid theory with a ﬁnite number of derivatives without any extra new degrees of freedom. But the definition of nonequilibrium ﬂuid variables is not ﬁxed and the theory is pathology-free only away from Landau (or Eckart) frame.

Given the situation, the current work indicates that there is always a tension between the three desirable features of hydrodynamics (i), (ii) and (iii), and to maintain causality and stability we have to give up at
least one of them. It is shown that the key features for causality and stability of the popularly existing (mentioned earlier) relativistic hydrodynamic theories can be translated into each other but it is never possible to have three of them simultaneously. It is further shown that no finitely truncated theory can be causal and an all-order resummation of gradient contributions is
equivalent to introducing new “nonfluid” degrees of freedom. The argument has been established with an exact derivation of relativistic hydro from a microscopic kinetic theory that explains the origin of the new degrees of freedom. The adopted metric signature is $g^{\mu\nu}=(-1,1,1,1)$ with natural units $c=k_B=\hbar=1$.

\section{Different hydro theories across a ‘fluid frame’ transformation}
We start with the BDNK energy-momentum tensor ($T^{\mu\nu}$) with conformal symmetry and no conserved charges as the following \cite{BDNK},
\begin{align}
&T^{\mu\nu}=(\varepsilon +{\cal{A}})\left[u^{\mu}u^{\nu}+\frac{\Delta^{\mu\nu}}{3}\right]+\left[u^{\mu}Q^{\nu}+u^{\nu}Q^{\mu}\right]-2\eta\sigma^{\mu\nu},
\nn
&{\cal{A}}={\chi}\left[3\frac{DT}{T}+\nabla_{\mu}u^{\mu}\right],~~
Q^{\mu}=\theta\left[\frac{\nabla^{\mu}T}{T}+Du^{\mu}\right].
 \label{BDNKeq}
 \end{align}
 The used notations read,
$D= u^{\mu}\partial_{\mu}$, $\nabla^{\mu}=\Delta^{\mu\nu}\partial_{\nu}$,
$\sigma^{\mu\nu}=\Delta^{\mu\nu}_{\alpha\beta} \partial^{\alpha}u^{\beta}$ with $\Delta^{\mu\nu\alpha\beta}=\frac{1}{2}\Delta^{\mu\alpha}\Delta^{\nu\beta}+\frac{1}{2}\Delta^{\mu\beta}\Delta^{\nu\alpha}
-\frac{1}{3}\Delta^{\mu\nu}\Delta^{\alpha\beta}$ and
$\Delta^{\mu\nu}=g^{\mu\nu}+u^{\mu}u^{\nu}$,
$\varepsilon=$energy density, $P=$pressure, $\eta=$ shear viscosity and $\chi,\theta$ are the first-order field correction coefficients. The BDNK formalism requires no additional degrees of freedom other than $T$ and $u^{\mu}$, but produces stable-causal modes only away from Landau frame that forgo the first principle definition of non-equilibrium fluid fields $T,u^{\mu}$. Here, we made an attempt to fix the deﬁnitions of ﬂuid velocity and the temperature in terms of the stress tensor. So we redeﬁne $u^{\mu}$ and $T$ in a way so that $T^{\mu\nu}$, expressed in terms of these redeﬁned ﬂuid variables, satisﬁes the Landau frame condition. The BDNK variables $u^\mu$ and $T$ relate the Landau frame velocity $\hat u^\mu$ and temperature $\hat T$ as,
$
u^\mu = \hat u^\mu +\delta u^\mu,~T=\hat T+\delta T$.
The shift variables $\delta u^\mu$ and $\delta T$ are small enough to be treated only linearly but include all orders of gradient corrections. The resulting frame transformed BDNK stress tensor in Landau frame is causal, uses only fluid fields $\hat{T},\hat{u}^{\mu}$ that are well defined by Landau frame choice, but it's gradient correction extends to infinity including all order derivative corrections. A theory with such infinitely many derivatives has practical limitations for simulation purposes. So it is recast in a finitely truncated series of equations by introducing new ‘non-ﬂuid’ variables just like the MIS theory in the following manner,
\begin{align}
 &\partial_\mu T^{\mu\nu}=0~,~~~~~~~~
 T^{\mu\nu}=\hat{\varepsilon}\left[\hat{u}^{\mu}\hat{u}^{\nu}+\frac{1}{3}\hat{\Delta}^{\mu\nu}\right]+\hat{\pi}^{\mu\nu},\nn
 &(1+\tilde\theta\hat D)\hat{\pi}^{\mu\nu}=-2\eta\hat{\sigma}^{\mu\nu}+\rho_1^{\mu\nu},\nn
 &(1+\tilde\chi\hat D)\rho^{\mu\nu}_1=(-2\eta)(-\tilde{\theta})\frac{1}{\hat{T}}\hat{\nabla}^{\langle\mu}\hat{\nabla}^{\nu\rangle} \hat{T}+\rho_2^{\mu\nu},\nn
 &(1+\tilde\theta\hat D)\rho^{\mu\nu}_2=(-2\eta)(-\tilde{\theta})\left(-\frac{\tilde{\chi}}{3}\right)\hat{\nabla}^{\langle\mu}\hat{\nabla}^{\nu\rangle} \hat{\nabla}\cdot\hat{u}+\rho_3^{\mu\nu},\nn
   &\vdots~~~~~~
 \label{MIS-type1}
\end{align}
The calculational details can be found in \cite{BMR}.
Eq.\eqref{MIS-type1} and so on construct a nested series of an infinite number of new degrees of freedom much in the same line as the conventional MIS theory given by,
$T^{\mu\nu}=\varepsilon(u^{\mu}u^{\nu}+\frac{1}{3}\Delta^{\mu\nu})+\pi^{\mu\nu},~(1+\tau_{\pi} D)\pi^{\mu\nu}=-2\eta\sigma^{\mu\nu}$, with $\tau_{\pi}$ as the relaxation time of shear viscous flow $\pi^{\mu\nu}$. So in order to have a finitely truncated theory and well defined fluid variables (in terms of Landau matching), condition (i) mentioned in introduction has to be compromised and new degrees of freedom are needed.

\section{Acausality in truncated Muller-Israel-Stewart-type theory}
In this study, we rewrite the MIS equations of motion without treating $\pi^{\mu\nu}$ as an independent degree of freedom and instead expressing it as a sum of order by order gradient corrections as,
$
\pi^{\mu\nu}=\sum_{n}\pi_n^{\mu\nu},~
\pi^{\mu\nu}_1=-2\eta\sigma^{\mu\nu},~
\pi_n^{\mu\nu}=-\tau_{\pi}D\pi^{\mu\nu}_{n-1}~,n \geq 2~$. This leads to,
$ \pi^{\mu\nu}=-2\eta\left\{\sum_{n=0}^{N}(-\tau_{\pi}D)^n\right\}\sigma^{\mu\nu}$, where upto $N^{\text{th}}$ order of temporal derivative corrections have been taken. For $N\rightarrow\infty$, the infinite sum results into a closed form to give the conventional MIS theory, $\pi^{\mu\nu}=-2\eta \left(1+\tau_{\pi}D\right)^{-1}\sigma^{\mu\nu}$.
It can now be rigorously proved that, for any finite value of $N$ (no matter how high it is) the theory violates the relativistic quantum theory causality condition,
\be
\text{Im}(\omega(k))\le |\text{Im}(k)|~.
\label{Heller}
\ee
Only an all order theory ($N\rightarrow\infty$) obeys Eq. \eqref{Heller} with the condition $\tau_{\pi}>\eta/(\epsilon+P)$. The details of the proof can be found in \cite{Mitra1}. So a truncated theory (no matter how high the gradient correction is) can never lead to a causal theory. Causality requires all order derivative corrections that is either folded as new degrees of freedom (MIS theory) or as field redefinition (BDNK theory).

\section{Microscopic origin of new degrees of freedom in hydrodynamics}
We start with microscopic kinetic theory in relaxation time ($\tau_{R}$) approximation that solves the out-of-equilibrium particle distribution function as,
\begin{align}
\tilde{p^{\mu}}\partial_{\mu}f=-\frac{\tilde{E}_p}{\tau_R}\delta f=-\frac{\tilde{E}_p}{\tau_R}\left(f-f_0\right)~,~~~~~~
 \delta f=-\frac{\left[\frac{\tau_R}{\tilde{E}_p}\tilde{p}^{\mu}\partial_{\mu}\right]f_0}
 {\left[1+\frac{\tau_R}{\tilde{E}_p}\tilde{p}^{\mu}\partial_{\mu}\right]}~.
 \end{align}
 Plugging it into the moment equation as
 $\frac{\pi^{\alpha\beta}}{T^2}=\int d\Gamma_p \tilde{p}^{\langle\mu} \tilde{p}^{\nu\rangle}\delta f $,
 we obtain an all order differential equation which can be recast in a truncated form as a set of equations with an in­finite number of new degrees of freedom as \cite{Mitra2},
\begin{align}
 \left(1+\tau_R D\right)\pi^{\alpha\beta}&=2\eta\sigma^{\alpha\beta} +\rho_1^{\alpha\beta}~,\nn
\left(1+\tau_R D\right)^2\rho_{1}^{\alpha\beta}&=-\eta\tau_R^2\left[\frac{2}{7}\nabla^2\sigma^{\alpha\beta}+\frac{12}{35}\nabla^{\langle\alpha}\nabla^{\nu}\sigma^{\beta\rangle}_{\nu}\right]+\rho_2^{
\alpha\beta}~,\nn
\left(1+\tau_R D\right)^2\rho_{2}^{\alpha\beta}&=\eta\tau_R^4\left[\frac{2}{21}\nabla^4\sigma^{\alpha\beta}+\frac{32}{105}\nabla^2\nabla^{\langle\alpha}\nabla^{\nu}\sigma^{\beta\rangle}_{\nu}+\cdots\right]\nn
\vdots
\label{dof}
\end{align}
The underlying microscopic descriptions of a hydrodynamic theory such as the kinetic theory are always pathology free. These causality related issues are basically the limitations of a particular coarse-graining method which is adjusted via the newer degrees of freedom. (for detailed discussion see \cite{Mitra2}).

\section{Conclusion and outlook}
In this work, several aspects of relativistic hydrodynamics in the context of causality and stability have been studied. The desirable features and limitations met by a stable-causal hydro theory have been discussed in detail with a possible microscopic origin as well. The results are quite interesting in finding a connection between different existing hydrodynamic theories and bridging a gap between them. The study opens several other aspects such as revealing the effect of field redefinition on the dispersion modes of the theory \cite{BMRS} and many more to explore in the future.

\section{Acknowledgements}
For the ﬁnancial support I acknowledge the Department of Atomic Energy, India.

\end{document}